\documentclass[11pt]{article}

\usepackage{amssymb}
\usepackage{amsmath}
\usepackage{graphicx}
\usepackage{subfig}
\usepackage[misc,geometry]{ifsym}
\usepackage{url}
\usepackage{hyperref}
\usepackage{lineno}
\usepackage{verbatim}
\usepackage{algorithm}
\usepackage{algpseudocode}
\usepackage{booktabs}
\usepackage{multirow}
\usepackage{balance}
\usepackage{verbatim}

\usepackage{setspace}
\newcommand{\Sol}[0]{{\cal X}}
\newcommand{\Neigh}[0]{{\cal N}}

\begin{document}

\title{Complex Networks Analysis of the Energy Landscape of the Low Autocorrelation Binary Sequences Problem}
\author{Marco Tomassini\\ 
Information Systems Institute, University of Lausanne, Switzerland}

\date{}
\maketitle%

\begin{abstract}
\noindent We provide an up-to-date view of the structure of the energy landscape of the
low autocorrelation binary sequences problem, a typical representative
of the $NP$-hard class. To study the landscape features of interest we use 
the local optima network methodology through exhaustive extraction of the optima
graphs for problem sizes up to $24$. Several metrics are used to characterize the
networks: number and type of optima, optima basins structure, degree and strength distributions,
shortests paths to the global optima, and random walk-based centrality of optima. 
Taken together, these metrics provide a quantitative and coherent explanation for
the difficulty of the low autocorrelation binary sequences problem
 and provide information that could be exploited by optimization heuristics for this problem, as well as 
 for a number of other problems having a similar configuration space structure. 

\end{abstract}

\section{Introduction}
\label{intro} 

Given a binary sequence $s = (s_1, s_2, \ldots, s_N)$ of length $N$ with $s _i \in \{-1,+1\}$, the autocorrelation
of $s$ is defined as:
$$
C_{k}(s) = \sum_{i=1}^{N-k} s_i s_{i+k} 
$$
The \textit{Energy} $E$ of the sequence $s$ is then given by:
$$
E(s) = \sum_{k=1}^{N-1} C_{k}^2(s),
$$
and the \textit{Merit Factor} $F(s)$ of $s$ has the form:
$$
F(s) = \frac{N^2}{2E(s)}
$$

Binary sequences of this type with minimal, or low, autocorrelation arise in several areas, especially in
telecommunications, but also in mathematics, statistical physics, and 
criptography. Although we shall not comment on these important theoretical and applied fields, limiting our
work to the combinatorial optimization aspects, the 
following references should be helpful~\cite{golay1982,marinari1994,jedwab2004}. 
The algorithmic goal is to minimize $E$ or, equivalently, to maximize $F$ and we shall call
the optimization problem the LABS (Low Autocorrelation Binary Sequences)
 problem for a given sequence $s$. The
problem becomes exponentially harder as $N$ increases since the number of admissible solutions grows
as $2^N$ and thus the time complexity in the worst case correspondingly grows as $O(2^N)$. In fact,
no algorithm other than complete enumeration or its variants, such as branch-and-bound, is known
for the problem, which therefore belongs to the $NP$-hard class.

LABS configurations enjoy a number of symmetries~\cite{packebusch2016low}. Among them, simply reversing or complementing a binary string leaves
the energy invariant. Because of these symmetries ground and higher states of the system are always degenerate,
i.e., there is more than one state with the same energy.
An important subset of sequences of odd length $N$ are called skew-symmetric if they satisfy 
$s_{j+l} = (-1)^l\:s_{j-l}$. Because only one half of the sequence is required, this reduces the search space size from $2^N$ to $2^{N/2}$ and thus makes the
problem less computationally intensive if minimum energy sequences are skew-symmetric~\cite{packebusch2016low}.

Global optimal configurations for $N$ up to $66$ have been found by exhaustive enumeration using
branch-and-bound techniques~\cite{packebusch2016low}. For larger $N$ a number of heuristic stochastic 
techniques have been used, including
simulated annealing, tabu search, and evolutionary algorithms  among others. Using these approaches that
cannot guarantee global optimality, best,
but not necessarily globally optimal solutions, have been found for many large odd $N$ values 
up to approximately $N=400$~\cite{bovskovic2017}. 
In fact, an empirical upper bound of about $12$ for $F(s)$ has been given by Golay~\cite{golay1982}, but for large $N$
there is a gap between the largest $F$ found and the bound~\cite{packebusch2016low} which means that  it is likely that
these current best sequences are not globally optimal.

The LABS problem has a strong link with spin systems in statistical mechanics. In fact,
LABS is formally similar to a long-range four-spin spin glass model~\cite{bernasconi} but it is completely
deterministic and has no quenched disorder. Because of this similarity, the model has been intensively studied with
the methods of statistical mechanics thereby producing many useful results (see, 
e.g.,~\cite{bernasconi,marinari1994,bouchaud1994}). The common features of these problems is
the presence of frustration, which causes the system to get stuck into suboptimal states
from which it is difficult to escape. However, we shall not pursue this direction further here. Instead,
in the present study we shall focus on an aspect of the LABS problem that has been less investigated: the
analysis of the structure of the energy landscape. This is very important to get a better understanding of
what type of heuristics are likely to be more efficient for searching the problem space, given that complete
enumeration is out of the question for $N$ sufficiently large. To our knowledge, there is only a previous study by
Ferreira et al.~\cite{ferreira2000} of the LABS energy landscape using the concept of barrier trees
and the notions of depth and difficulty. 
Our goal in the present contribution is to deepen our understanding of the energy landscape of the LABS problem
beyond that offered by~\cite{ferreira2000} and, at the same time, uncover a number of features that are common to
most hard combinatorial optimization problems that have fitness landscapes similar to the LABS landscape.
In order to investigate the global structure of the energy landscape
 we shall use an approach based on
Local Optima Networks (LONs) previously developed in our group~\cite{pre08} in which the  $2^N$ 
original confi\-gurations are
reduced to those that are local optima and to their connexions only. It has been shown in a number of works that
the LON approach is a valuable one and, by studying the corresponding graph with the methods of complex networks,
 it can give useful insights on the factors causing the hardness of many combinatorial
problems, see e.g.,~\cite{pre08,daolio2011communities}. 

The rest of the article is structured as follows. For the sake of self-containedness, in the next section we present the main 
ideas of Local Optima
Networks (LONs). This is followed by the analysis of the exhaustive LONs generated by LABS instances of size $N$ for 
$N$ up to $24$. The results are then discussed in terms of the relationship between fitness landscape structure
and problem difficulty. Finally, we present our conclusions.

\section{Fitness Landscapes and Local Optima Networks}
\label{lons}

Let us consider a discrete problem $ \mathcal P$ and an instance $\mathcal I_{\mathcal P}$ of $ \mathcal P$. 
$I_{\mathcal P}$'s \textit{fitness landscape}~\cite{reidys2002combinatorial} is a triplet $(\Sol, \Neigh, f)$ where $\Sol$ is a set of admissible solutions, \textit{i.e.}, a search space; $\Neigh : \Sol \longrightarrow 2^\Sol$, a neighborhood structure;
that is, a function that assigns to every $x \in \Sol$ a set of neighbors $\Neigh(x)$, and $f : \Sol \longrightarrow \mathbb{R}$ is a fitness function that provides the fitness, or objective value, of each $x \in \Sol$. In the LABS case fitness is sinonymous with the energy of a given binary sequence and the set of admissible solutions $\Sol$ is the set of all possible binary sequences
of length $N$ formed with the symbols $\{-1, 1\}$. The neighborhood $\Neigh(x)$ of a solution $x$
can be defined in various ways. Here we use the standard 1-bit flip operator that changes the sign of single digit in the sequence $x$ so
that $|\Neigh(x)| = N$. The elements $x$ of $\Sol$ are related to the ordinary Boolean hypercube with elements $b_i \in \{0,1\}$ by 
the transformation $x_i = 2b_i - 1$.

The LON idea~\cite{pre08,tec11,ea11} starts from the above concepts and builds a new graph $G_w=(V,E)$ in which $V$ is the
set of vertices which are \textit{local optima} in the given fitness landscape and $E$ is the set of arcs joining
two given vertices when they are directly reachable from each other in a sense that will be made explicit below.
We now explain the construction of the graph $G_w$ in more detail.

%Our study considers two search spaces or solution representations: binary strings (NK landscapes) and permutations (QAP). For each case, we selected the most basic neighborhood structure, as described in Table \ref{tab:moves}. The single bit-flip operation changes a single bit in a given binary string, whereas the pairwise exchange operation exchanges any two positions in a permutation, thus transforming it into another permutation.

%\begin{table}[!ht]
%\begin{center}
% \caption{Search space and neighborhood structure characteristics. Where $N$ represents the genotype length. \label{tab:moves}}
%\begin{tabular}{lcc}
%\toprule
%   &  Binary &  Permutation \\
%   %\hline
%   \cmidrule{2-3}
%    Search space size &  $2^N$ &  $N!$ \\
%    Neighborhood &  single bit-flip &  pairwise exchange \\
%    Neighborhood size &  $N$ &  $N(N-1)/2$ \\
%%\hline
%\bottomrule
%\end{tabular}
%\end{center}
%\end{table}

The nodes $V$ in the network are  local optima (LOs) in the search space. 
For a minimization problem, a solution $x \in \Sol$ is a local optimum iff $\forall x^{\prime} \in \Neigh(x)$, $f(x^{\prime}) \geqslant f(x)$.
For a maximization problem the inequality is reversed.
%Notice that in this work we do not target specifically neutral fitness landscape with large plateaus. However, this definition of local optima is still relevant for small amounts of neutrality. For fitness landscape with high levels of neutrality, please refer to the definitions of previous work \cite{tec11} where the nodes are local optima plateaus.
LOs are extracted using a best-improvement descent local search, as given in Algorithm~\ref{alg:HC} in which
%which defines a mapping from the search space $\Sol$ to the set of locally optimal solutions $\Sol^*$. 
when selecting the fittest neighbor (line 4), ties are broken at random. 
%\ntext{In line 4 of procedure \ref{alg:HC}, when several neighboring solutions reach the best fitness of the neighborhood, one random solution among the best ones is set.} 

\begin{algorithm}[H]
\begin{algorithmic}[1]
\Procedure{Local Descent}{}
   \State $x \gets $ random initial solution
   \While{$x \not=$ Local Optimum}
      \State set $x^\prime \in \Neigh(x)$, s.t. $f(x^\prime) = min_{y \in \Neigh(x)} f(y)$
     \If {$f(x) \le f(x^\prime)$}
               \State $x \gets x^\prime$
       \EndIf
   \EndWhile
\EndProcedure
\end{algorithmic}
\caption{Best-improvement local descent (bild)\label{alg:HC}}
\end{algorithm}

The edges $E$ in the network are defined according to a distance function $dist$ and a positive integer $D > 0$. The distance function represents the minimal number of moves between two solutions by a given search (mutation) operator which is a one-bit flip in our case. There is an edge $e_{ij}$ between local optima $LO_i$ and $LO_j$ if a solution $x$ exists such that $dist(x, LO_i) \leqslant D$ and $bild(x) = LO_j$. In other words, if $LO_j$ can be reached after mutating $LO_i$ and running best improvement local descent from the mutated solution. The weight $\tilde{w}_{ij}$ of this edge is $\tilde{w}_{ij}= \sharp \{ x \in \Sol ~|~ dist(x, LO_i) \leqslant D \mbox{ and } bild(x) = LO_j \}$. That is, the number of $LO_i$ mutations that reach $LO_j$ after local descent. This weight can be normalized by the total number of solutions, $\sharp \{ x \in \Sol ~|~ dist(x, LO_i) \leqslant D\}$, within reach at distance $D$: $w_{ij}=\tilde{w}_{ij} / \sum_{j} \tilde{w}_{ij}$.

Summarizing, the LON is the weighted graph $G_w=(V,E)$ where the vertices $v_i \in V$ are the local optima, and there is an edge $e_{ij} \in E$, with  weight $w_{ij}$, between two vertices $v_i$ and $v_j$ if $w_{ij} > 0$.
According to the definition of weights, $w_{ij}$ may be different than $w_{ji}$. Thus, two weights are needed in general, and $G_w$ is a weighted, directed transition graph.

The above systematic construction is only possible for problem instances of small to medium size. For larger instances
it is possible to resort to reliable sampling techniques, for instance as described in~\cite{thomson2019}. In the present 
work the problem sizes considered have allowed us to use exhaustive enumeration.

To give an idea of how a LON looks like, the following Fig.~\ref{lon} depicts the LABS LON for $N=8$. For $N > 10$ the graphs become
too dense and direct visualization is not useful. By the way, we note that
the ratio of the total number of optima to the number of global optima in this small LON is $1.5$ but it quickly increases 
with increasing $N$ (see Sect.~\ref{avdeg}). 
To analyze the network structure in the general case it is better to resort to relevant graph metrics that
summarize important graph features in a single number, a few numbers, or a probability distribution. This is the  goal of the next section.
\begin{figure*}[h!]
  \begin{center}
   \includegraphics[width=0.7\textwidth]{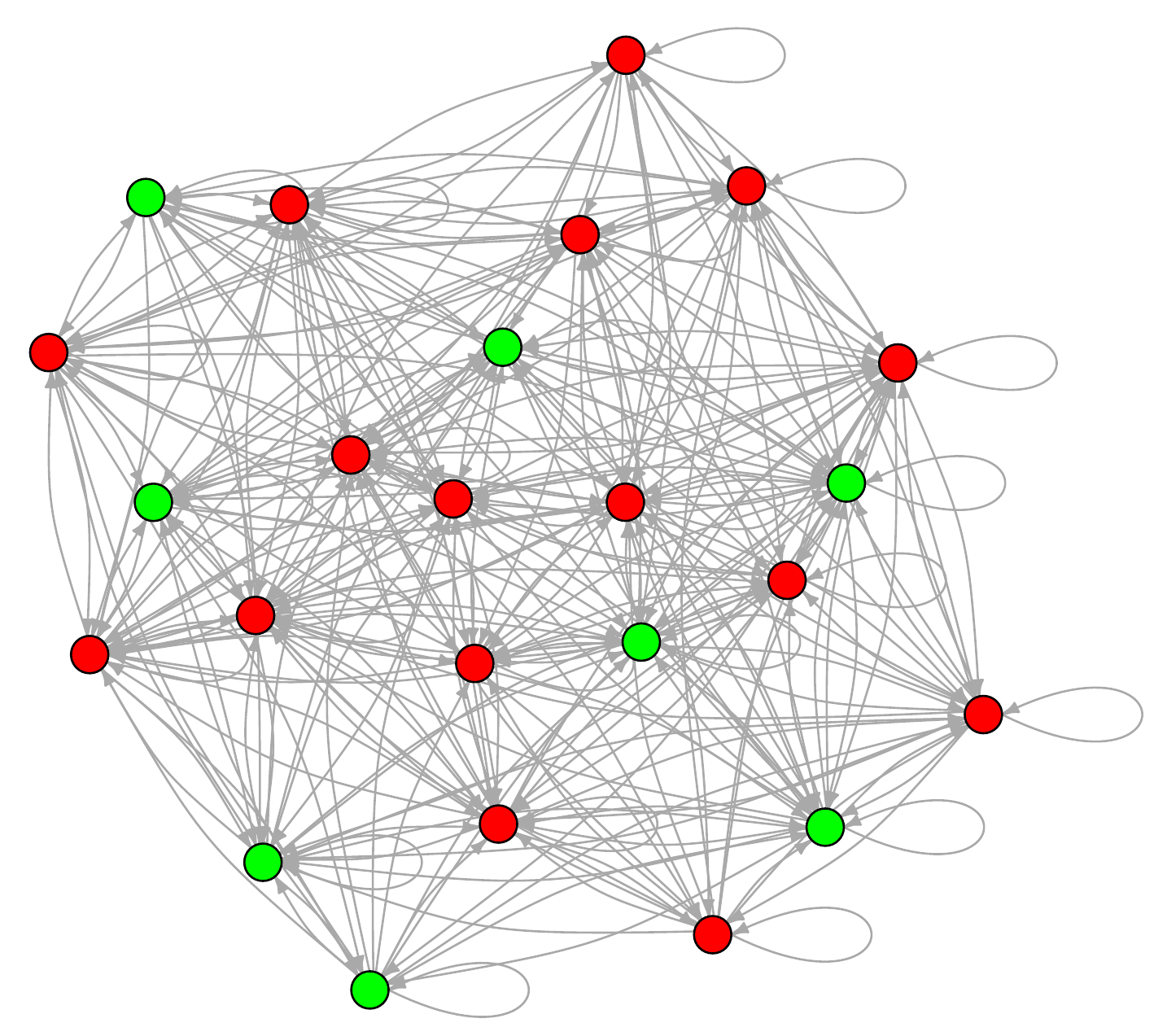}  
   \end{center}
  \caption{The Local Optima Network of the LABS problem with $N=8$. 
  Directed weighted edges stand for transitions frequencies between the basins corresponding to the minima nodes. Self-loops
  represent the fraction of transitions that remain in the source basin.
  Here edge weights are drawn equal for the sake of clarity. Global minima are in red. The other local optima are in green.}
\label{lon}
\end{figure*}
We also mention that LONS can be seen as the discrete equivalent of the \textit{inherent structures} of 
the free or potential energy hypersurfaces in the continuous parameter spaces of chemical-physical systems 
such as macromolecules and atom clusters~\cite{inherent}.

\section{LON Metrics}

Once we have built a LON, we are in a position of using tools from the science of complex
networks to analyze the corresponding graphs. There exist many metrics for describing complex networks. Among them, the following ones are those that have proved
useful in characterizing the difficulty of problem instances through their corresponding LONs in previous
 work~\cite{pre08,tec11}. They are briefly defined here for the sake of self-containedness, more details can be found, for example, in the books~\cite{newman2018,barabasi2016}. 

\paragraph{\bf Network size.}
The number of vertices in the LON and its variation with $N$ gives information about
the difficulty for a searcher to find the global optimum. It also provides data on the
sizes of the associated attraction basins.

\paragraph{\bf Strength.} 
This term refers to the generalization of the vertex degree to weighted networks. 
It is defined as the sum $s_i$ of  weights of the edges from node $i$ to its neighbors $\Neigh(i)$,
$$
s_i = \sum_{j \in \Neigh(i)} w_{ij},
$$
where $w_{ij}$ is the weight of the edge connecting nodes $i$ and $j$. For directed graphs
one speaks of outgoing strength which is computed on the edges outgoing from the node while
incoming strength refers to incoming edges.

\paragraph{\bf Degree, strength, and edge weight distribution functions.}
These discrete distributions give, respectively, 
the frequency of a given node degree, node strength, or edge weight in the network. 
These distributions are useful for evaluating whether they are, for instance, 
homogeneous or heterogeneous, unimodal or multimodal.

\begin{comment}
\paragraph{\bf Disparity.} 
A given value of a node's strength can be obtained with very different values of edge weights.
The contributing weights could be of about the same size or they could be very different. 
To measure the degree of heterogeneity of a node's edges \textit{disparity} can be used. 
It is defined as follows:
$$
Y_{2}(i) = \sum_{j \in \Neigh(i)} \left( \frac{w_{ij}}{s_i} \right)^2.
$$
If all the connections are of the same order then $Y_2$ is small
and of order $1/k$ where $k$ is the vertex degree. 
On the other hand,
if there is a small number of high weight connections $Y_2$ is larger and may approach unity.
\end{comment}

\paragraph{\bf Average shortest paths.} 

The average value $\langle L \rangle $  of all two-point shortest paths in a graph give an idea of the typical distances between nodes.
It is given by:

$$
\langle L \rangle =   \frac{2}{N(N-1)}\sum _{i=1}^N \sum _{j > i} l_{ij}.
\label{eq:av-path-len}
$$

\noindent Where $l_{ij}$ is the shortest path between vertices $i$ and $j$.
In LON networks edges are directed and weighted and thus Dijkstra's algorithm is usually employed to compute
them. These paths always exist because complete LONs are strongly connected.

\paragraph{\bf Network centrality.}
Centrality measures of various kinds are often used, especially in social networks, to characterize
actors that are centrally located in some specified sense. Centrality can also be used in other contexts and we shall use PageRank here to investigate the centrality of optima in a LON graph.

\begin{comment}

\paragraph{\bf Clustering coefficient.}

The clustering coefficient $C(i)$ of a node $i$  is defined as the ratio between the $e$ edges 
that actually exist between the $k$ neighbors of $i$ and the number of possible edges between these nodes:
$$
C(i) = \frac{e}{{{k} \choose{2}}} = \frac{2e}{k(k-1)}.
\label{eq:clust-coeff}
$$
The clustering coefficient can be interpreted intuitively as
the likelihood that two of node $i$'s neighbors are also neighbors.
The \textit{average clustering coefficient} $\bar C $ is the average 
of $C(i)$ over all $N$ vertices in the graph $G$,
$i \in V(G)$: $\bar C  = (1/N )\sum _{i=1}^N C(i).$  

\end{comment}

\section{Results}
\label{resul}

In this section we show and discuss the main results obtained by constructing and analyzing the full LONs for LABS
problems with size $N$ in the range $3$ to $24$. We remark that, in contrast with classical combinatorial problems
such as QAP or TSP, or spin glasses with quenched disorder in which many different instances can be generated for a given size $N$, LABS,
being completely deterministic, has only one instance for each problem size $N$ with the vertices of the
$N$-ary hypercube $\{-1,1\}^N$ as admissible solutions. This feature does not allow to
take averages over a sample of instances as in the other cases.

\subsection{LON Sizes}

\begin{figure*}[h!]
  \begin{center}
   \includegraphics[width=0.49\textwidth]{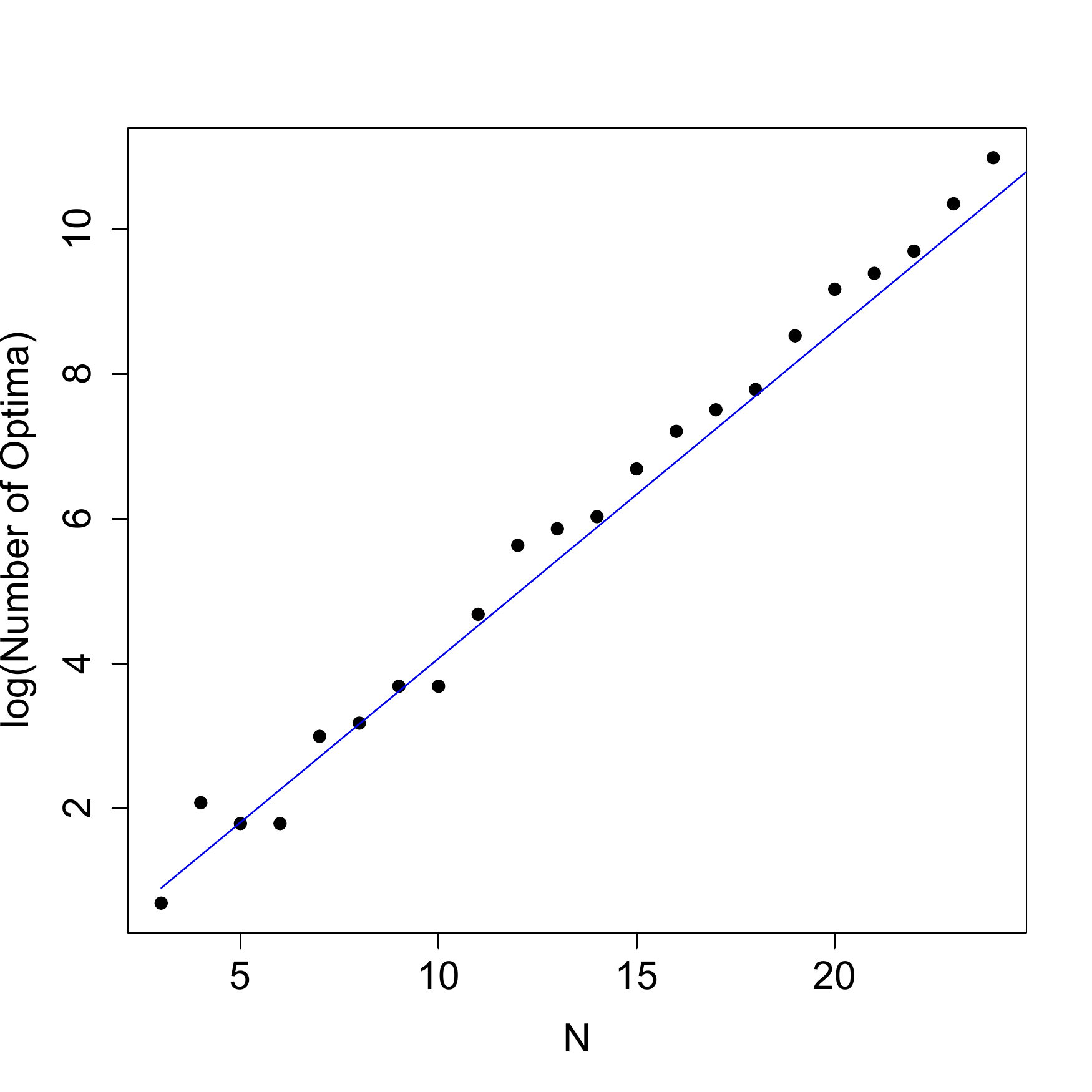}  
      \includegraphics[width=0.49\textwidth]{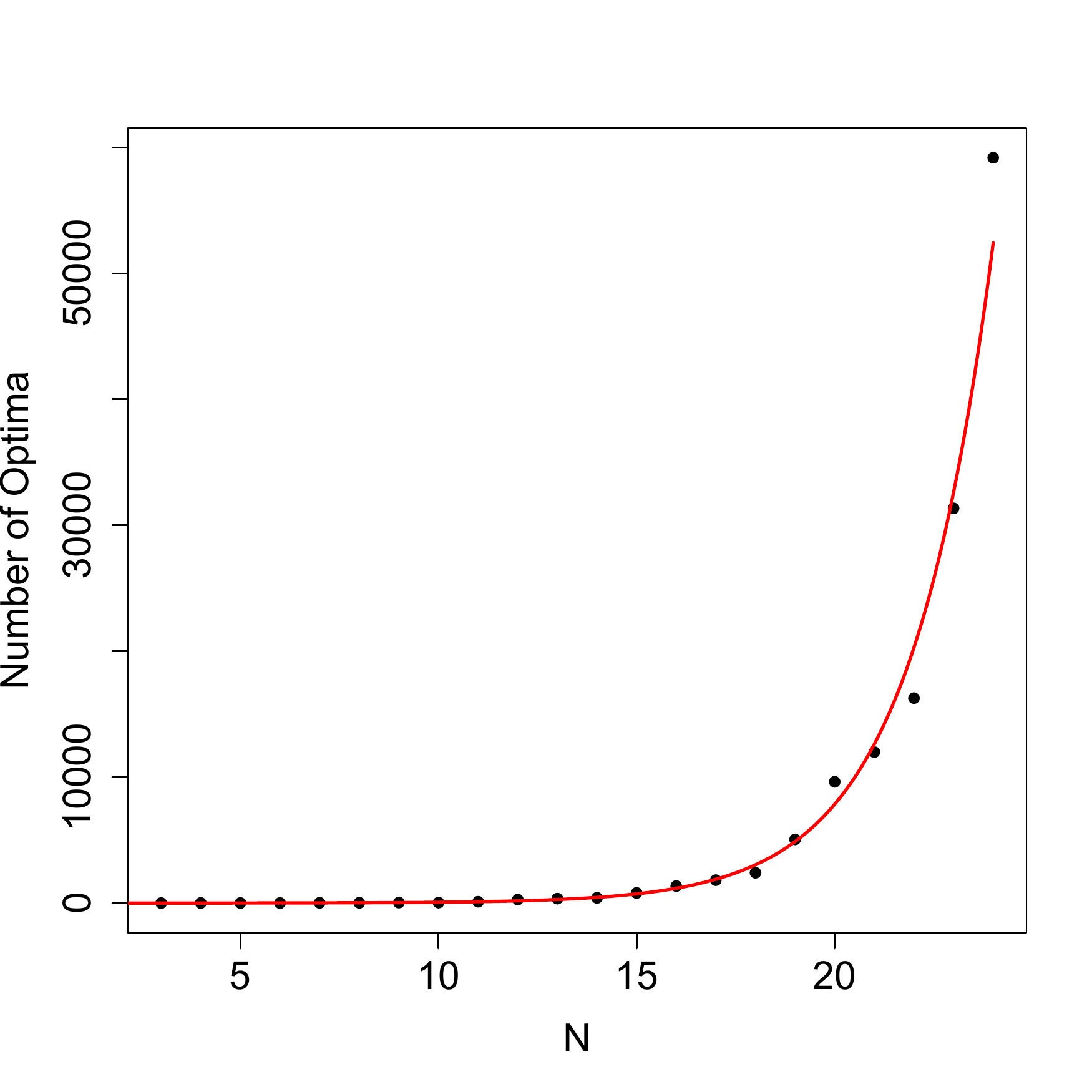}
   \end{center}
  \caption{Number of optima in LABS problems as a function of the problem size $N$.
  Left Image: lin-log scale with a linear fit. Right image: linear scale with an exponential curve fit.
  Residual std. error: 0.2852, $R^2= 0.9915$, p-value: $<$ 2.2e-16. The least squares regression line
  on the left image has equation: $y = 0.47459 \: x - 0.5233$.}
\label{sizes-log}
\end{figure*}

Figure~\ref{sizes-log}, left image, shows the number of vertices in the LON with growing $N$ on a semi-log
plot. The number of optima
grows exponentially with the problem size as in similar hard problems, e.g., highly 
epistatic $NK$-landscapes~\cite{kauffman93}, quadratic assignement problems~\cite{tayarani2015qap}, and the number partitioning problem~\cite{ferreira1998npp}
among others. The equivalent right image, with an exponential fit of the data on linear scales, shows
perhaps more clearly the rapid increase of the number of optima.
Hard instances of this class of problems all have very rugged cost landscapes, making
the search for the global optimum difficult.

\begin{figure*}[h!]
  \begin{center}
   \includegraphics[width=0.6\textwidth]{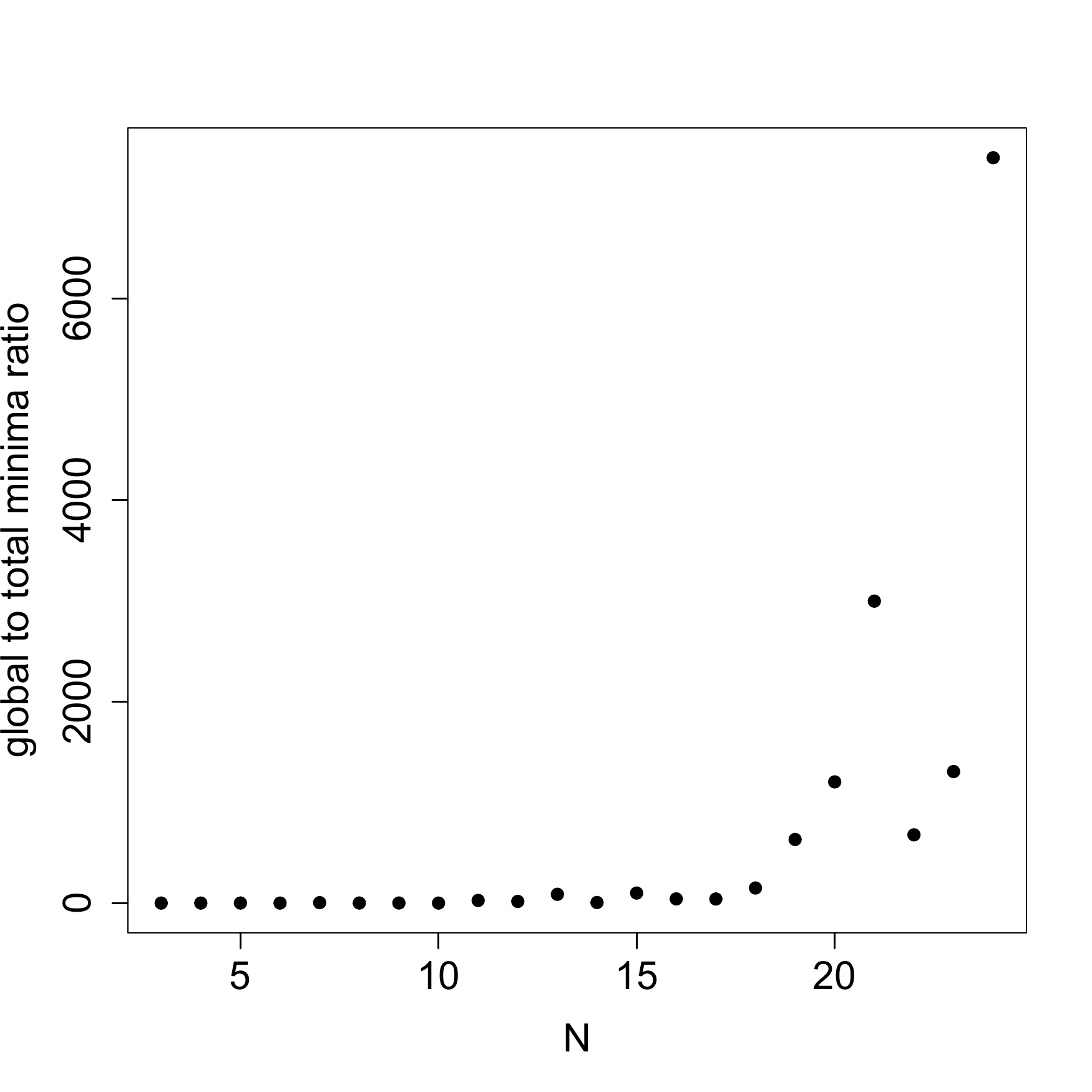}  
   \end{center}
  \caption{The plot represents the ratio of the total number of optima to the number of global optima in LABS problems as a function of the problem size $N$. For $N$ moderately large, i.e., around $N=19$ and larger the number of optima that are global minima becomes a small
  fraction of the total number of optima in the energy landscape.}
\label{ratio}
\end{figure*}

Since the global minimum always has some amount of degeneracy due to symmetry operation
equivalencies, it is also of interest to compute the ratio of the total number of optima 
 to the number of equivalent global optima for each $N$. This is shown in Fig.~\ref{ratio}.
From the plot it is clear that this ratio stays low for small $N$ but, as soon as $N$ nears
$20$ the ratio takes off exponentially, although there are some fluctuations due to differences
in symmetry. This behavior shows that the search for the global minimum remains
exponentially harder for increasing $N$ despite the fact that there are several global minima.

\subsection{Average Degree}
\label{avdeg}

LABS LONs are weighted but it is still informative to plot their mean degree $\langle k \rangle$ by ignoring directions and weights
and taking only the number of links into account. This gives Fig.~\ref{avdeg} in which we see that the LONs are rather
densely connected and $\langle k \rangle \propto N$.

\begin{figure*}[h!]
  \begin{center}
   \includegraphics[width=0.6\textwidth]{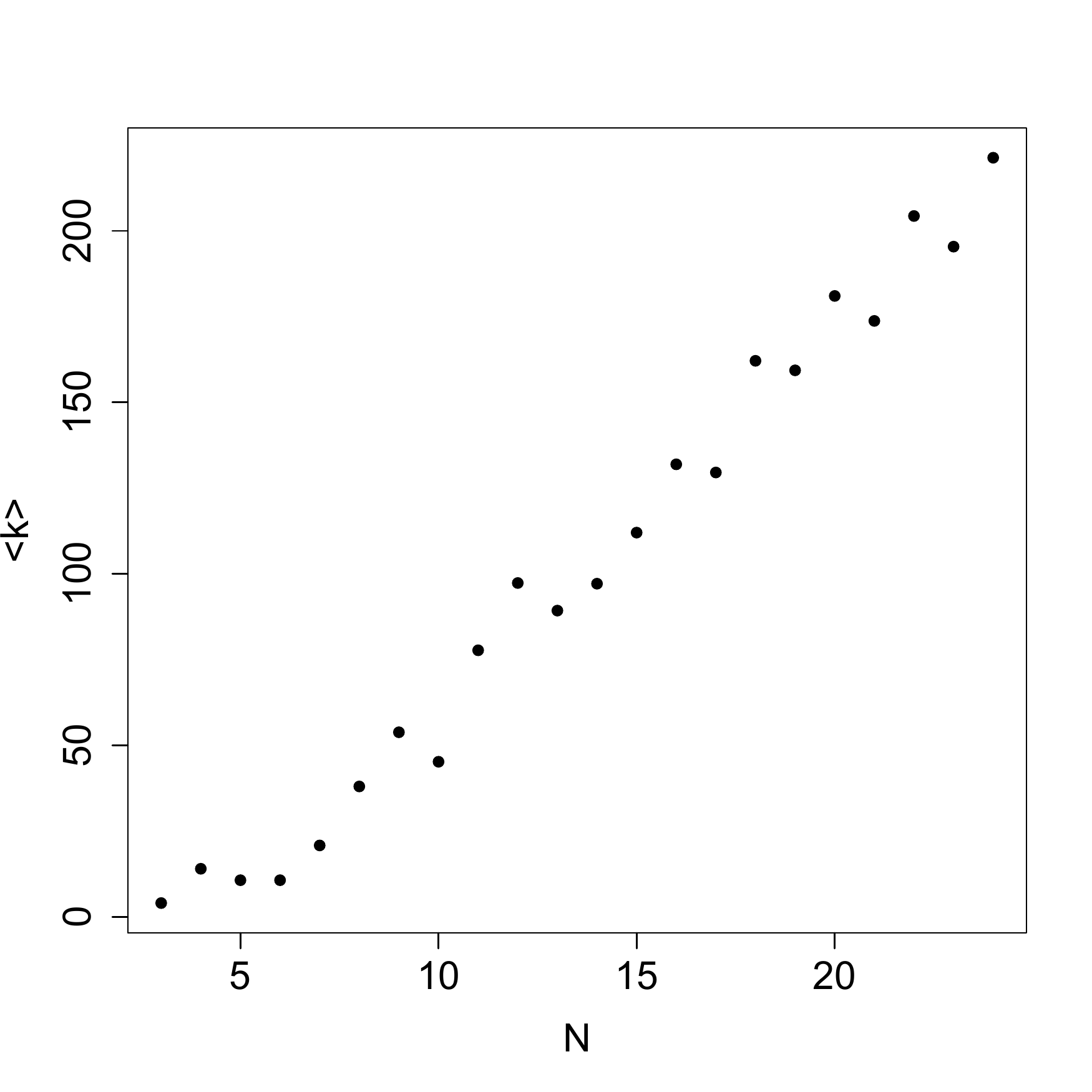}  
   \end{center}
  \caption{Average degree of the LABS LONS as a function of problem size $N$. Edge weights and direction are
  not taken into account for the computation, i.e. LONS are considered undirected and unweighted.}
\label{avdeg}
\end{figure*}

\subsection{Average Strength}

\begin{figure*}[h!]
  \begin{center}
   \includegraphics[width=0.45\textwidth]{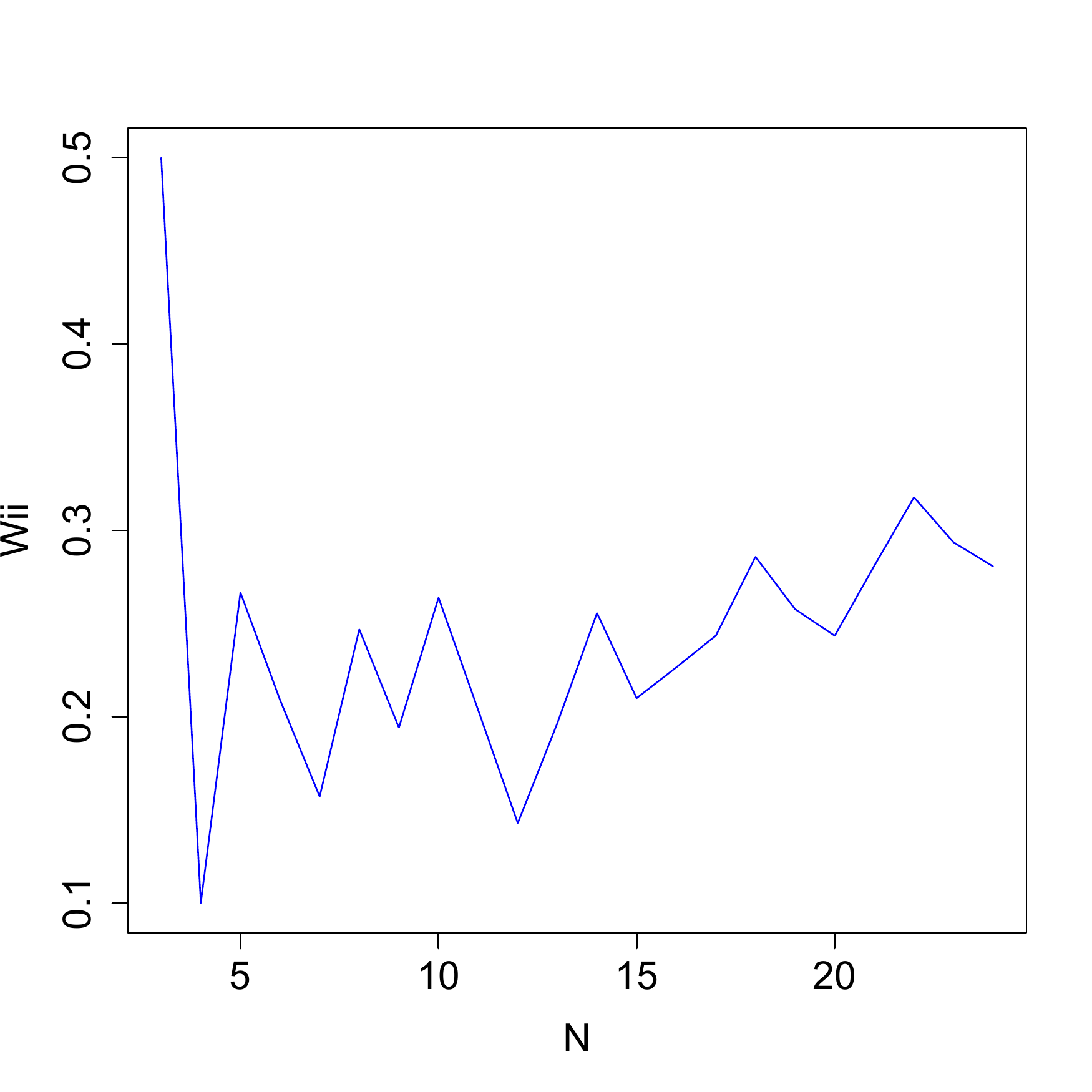}  \hspace{1cm}
   \includegraphics[width=0.45\textwidth]{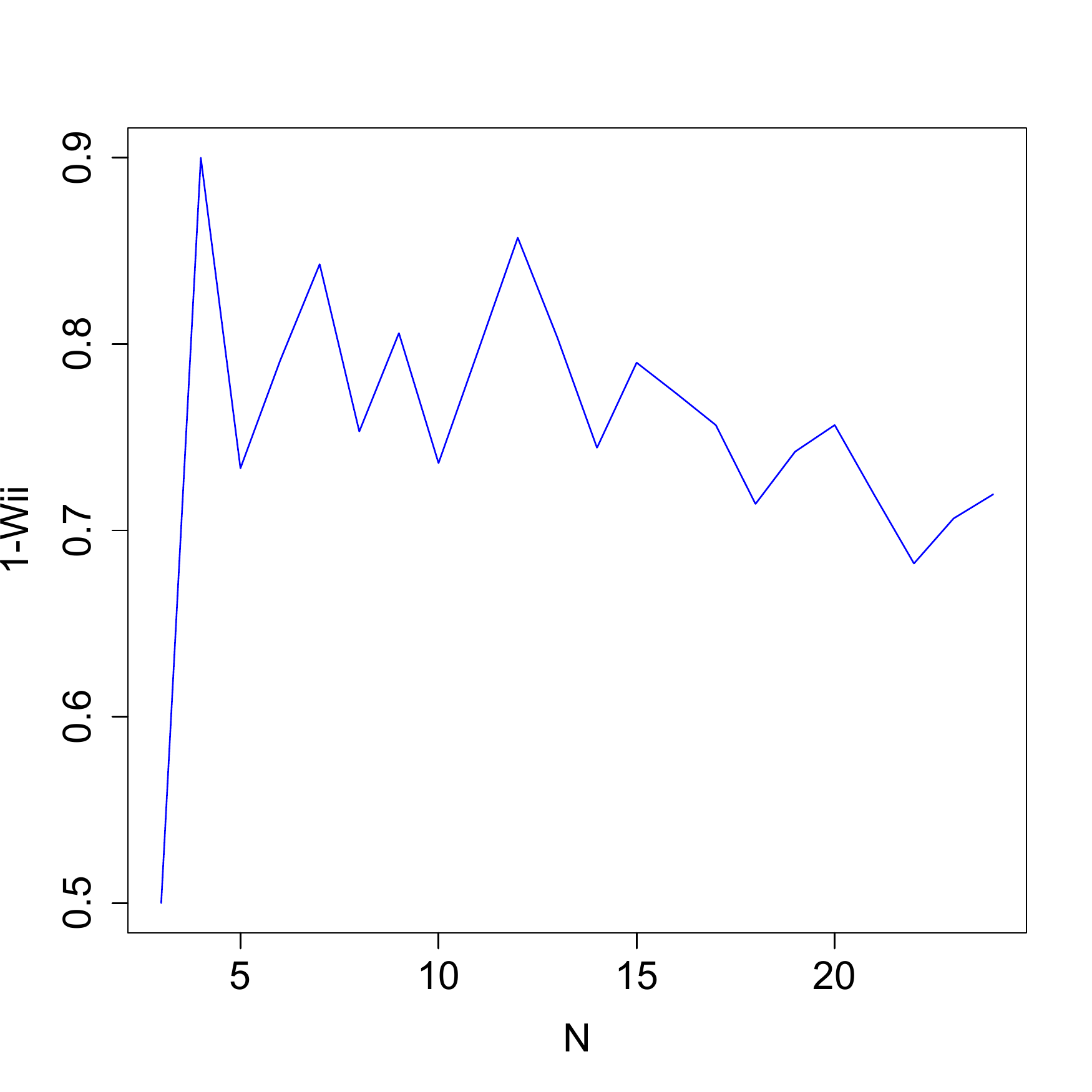} 
   \end{center}
  \caption{Average strength in LON graphs as a function of LON size $N$.
  Average link strength for self-loops $w_{ii}$ (left image) and outgoing links
  $1-w_{ii}$ (right image).}
\label{strength}
\end{figure*}

Node strength is informative because it is related to the probability of transition between
energy basins. As such, it can provide useful information on the likely behavior of 
optimization methods that search the energy landscape.
Figure~\ref{strength} (left image) depicts the average strength $\langle w_{ii} \rangle$, i.e. the average weight of the transitions from a given
optimum to itself. This value is a proxy for the average probability of remaining in a given energy basin once it has been reached.
This empirical probability increases with increasing $N$ showing that a search that has reached a suboptimal 
minimum has an increasing tendency to remain in its basin.

Since strength values are normalized to one, $w_{ij} = 1 - w_{ii}$ with $i \ne j$. Figure~\ref{strength} (right image) shows 
the average strength of all the other outgoing links. Clearly, this view is the complementary of the previous one
and it shows that outgoing, i.e., interbasin transitions become less probable with increasing $N$.

\subsection{Degree Distribution Functions}

\begin{figure*}[h!]
  \begin{center}
   \includegraphics[width=0.495\textwidth]{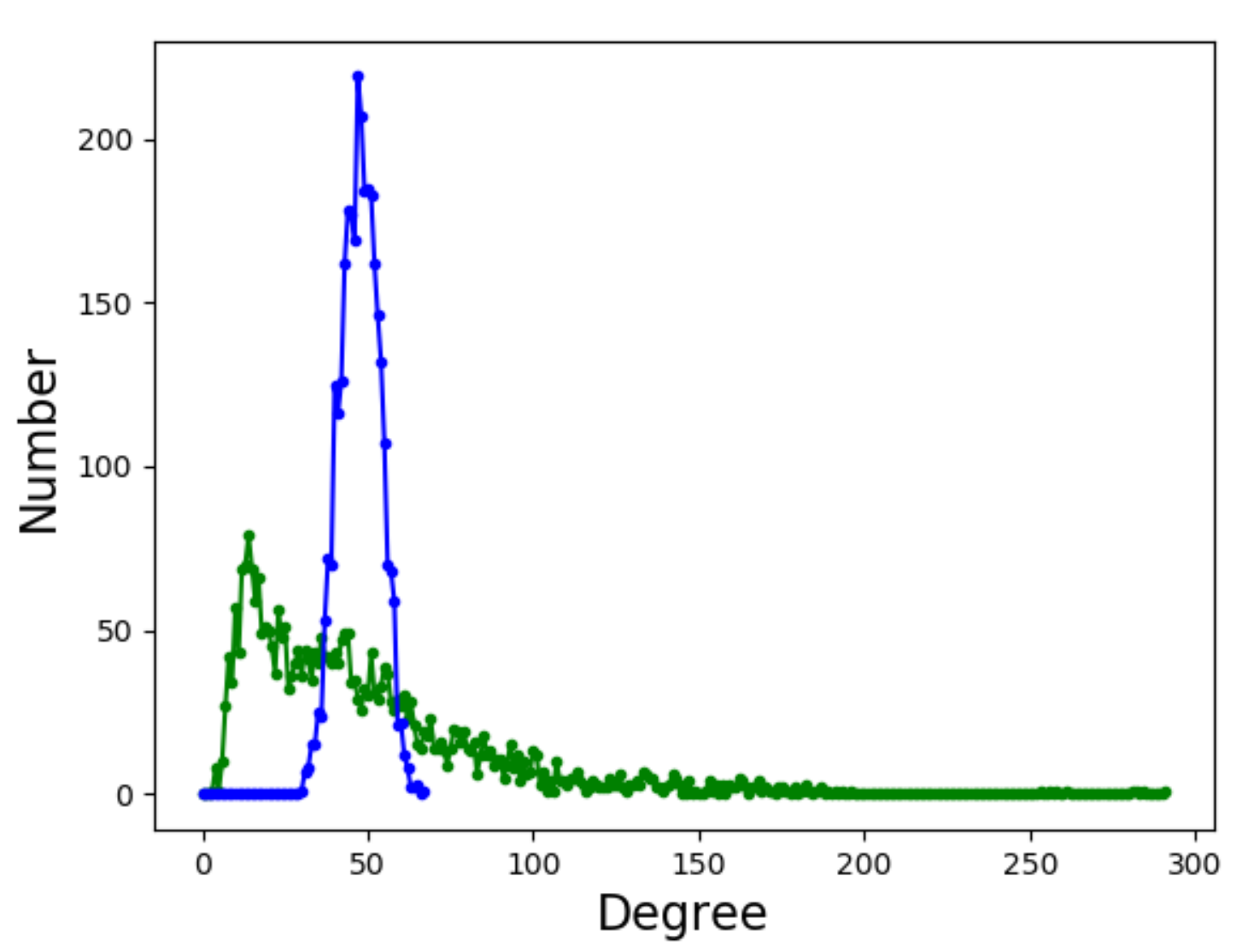}  
   \includegraphics[width=0.495\textwidth]{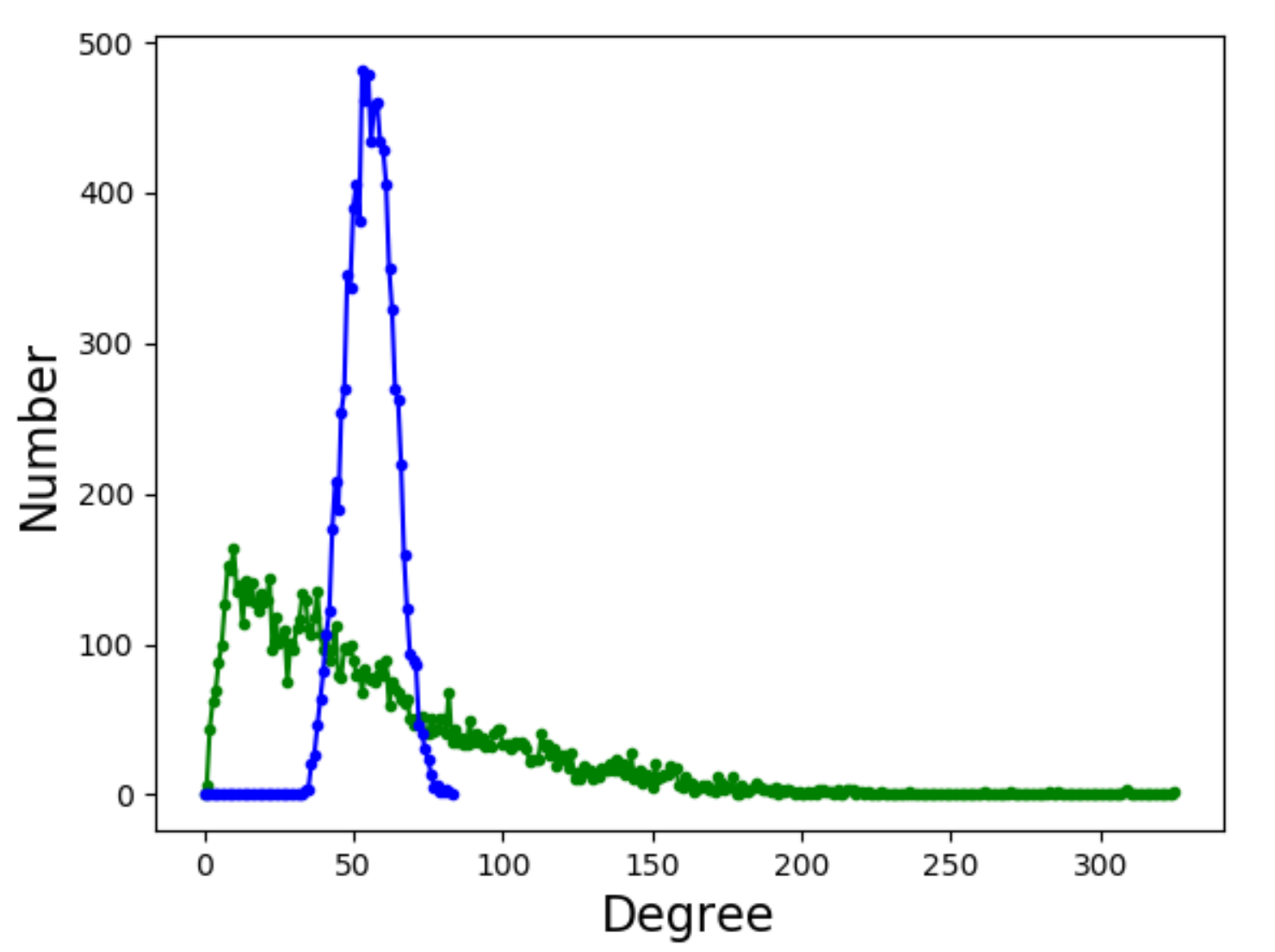}  
   \end{center}
   \vspace{-0.4cm}
      \footnotesize
    \hspace{2.8cm}$N=17   \hspace{5.2cm} N=20$
  \caption{Degree distribution function for incoming (green curve) and outgoing (blue curve) links for $N=17$
  (left plot) and $N=20$ (right plot).}
\label{dd}
\end{figure*}

The degree distribution function of a complex network may give some useful indication about the 
general structure of the latter, for example whether it belongs to a known class of distributions such
as Poissonian, exponential, or power law. Here we deal with directed graphs and thus we have
two distinct degree distributions: the incoming links distribution and the outgoing links distribution.
We found that the LON degree distributions all show the same trend for different $N$ values except for
very low $N$ for which the graphs have too few vertices.
To illustrate the typical tendency, we show the in and out distributions for $N=17$ and  $N=20$ in 
Fig.~\ref{dd}.
 
The remarkable thing is that in and out distributions are very different. We did not try to fit any
particular function to these empirical distributions; however, in a qualitative manner the outgoing links
distribution (blue curve) appears to be close to Poissonian with a rather narrow peak and a smaller 
fluctuation around the mean. The incoming
links distribution (green curve), on the other hand, is broader and has a much longer tail indicating that there is a non-negligible 
number of LON vertices with a large number of incoming links. This suggest that those nodes
could play a role in dynamical processes on LONs. We shall see later that other
measures tend to confirm this idea. For weighted networks strength distributions can also be computed
but since they don't bring new information in this case, we refrained from showing them to save space.

\subsection{Basin Sizes}
\begin{figure*}[h!]
  \begin{center}
   \includegraphics[width=0.6\textwidth]{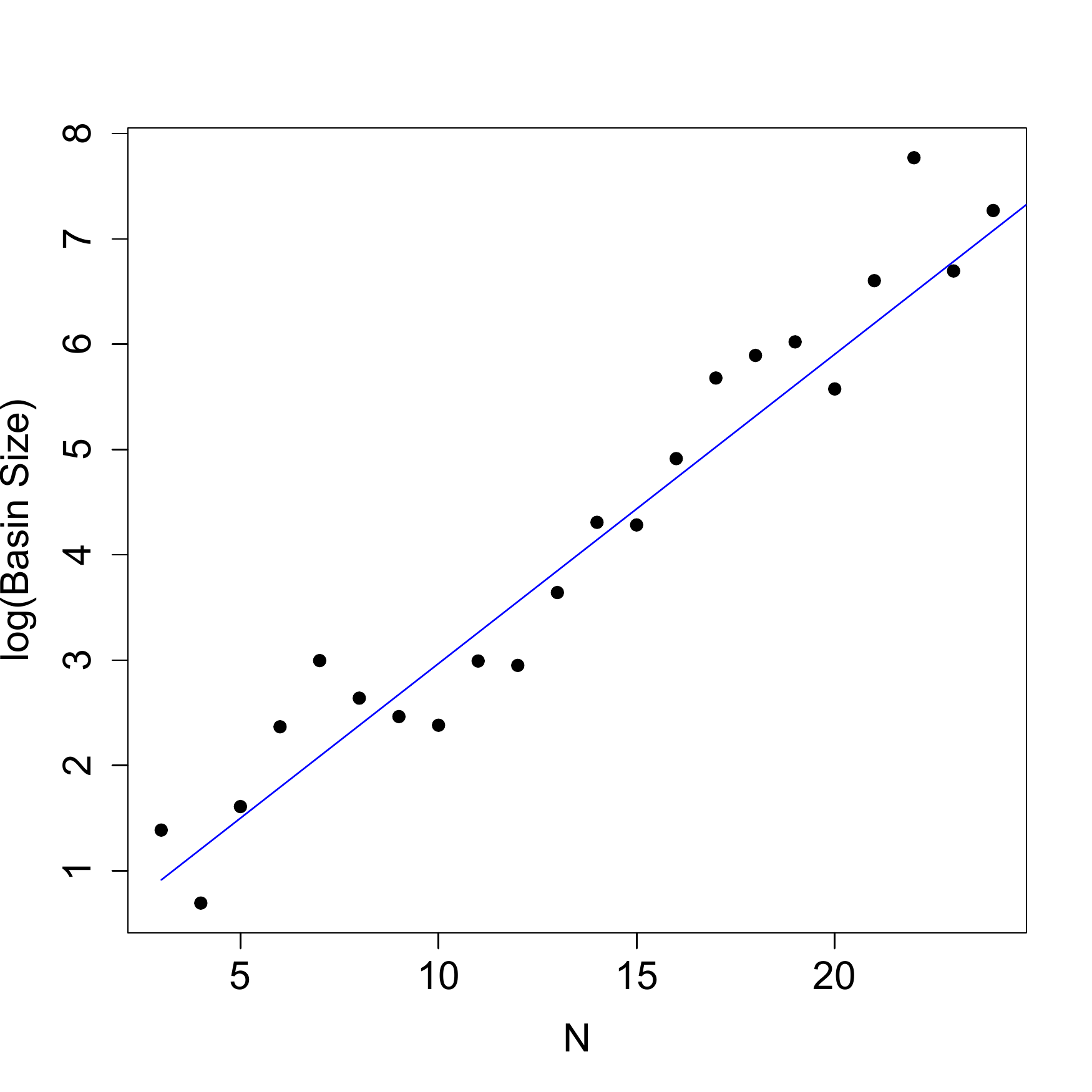}  
   \end{center}
  \caption{Basin Sizes of the global minimum as a function of $N$ on lin-log scale.
  A linear fit (blue line) gives a
  residual sd. err.: $0.4956$, $R^2=0.9419$, p-value: 4.895e-14. The equation of the least 
  squares regression line is: $y = 0.30765\: x - 0.01078$.}
\label{basins}
\end{figure*}

A useful by-product of the exhaustive LONs construction is that each admissible solution can be assigned to
the basin of attraction of one of the optima in the LON. This allows us to define a basin's size as
being equal to the number of configurations that end up reaching the corresponding optimum using 1-bit flip
and best improvement.
Figure~\ref{basins} shows the size of the global optimum basin as a function of $N$.

First, let us observe again that all global energy minima are degenerate, i.e., there
is always more than one global minimum with the same energy. For the figure, we simply plotted the
size of one of the global minima chosen at random since all of them have the same or similar size.
We observe that the size of the basin corresponding to a global optimum increases 
roughly exponentially with $N$. This
would seem to suggest that it should be easier to find during an heuristic search. However, this effect
is more than compensated for by the fact that the number of optima also increases exponentially with $N$ as shown in Fig.~\ref{sizes-log}.
In this highly rugged landscape, searches are thus likely to get stuck into one of the many good but not globally optimal minima. 
This phenomenon has previously been observed in other well known hard combinatorial optimization 
problems such as $NK$ landscapes and QAP~\cite{pre08,tayarani2015qap,daolio2010local,hernando2019}.

\begin{figure*}[h!]
  \begin{center}
   \includegraphics[width=0.495\textwidth]{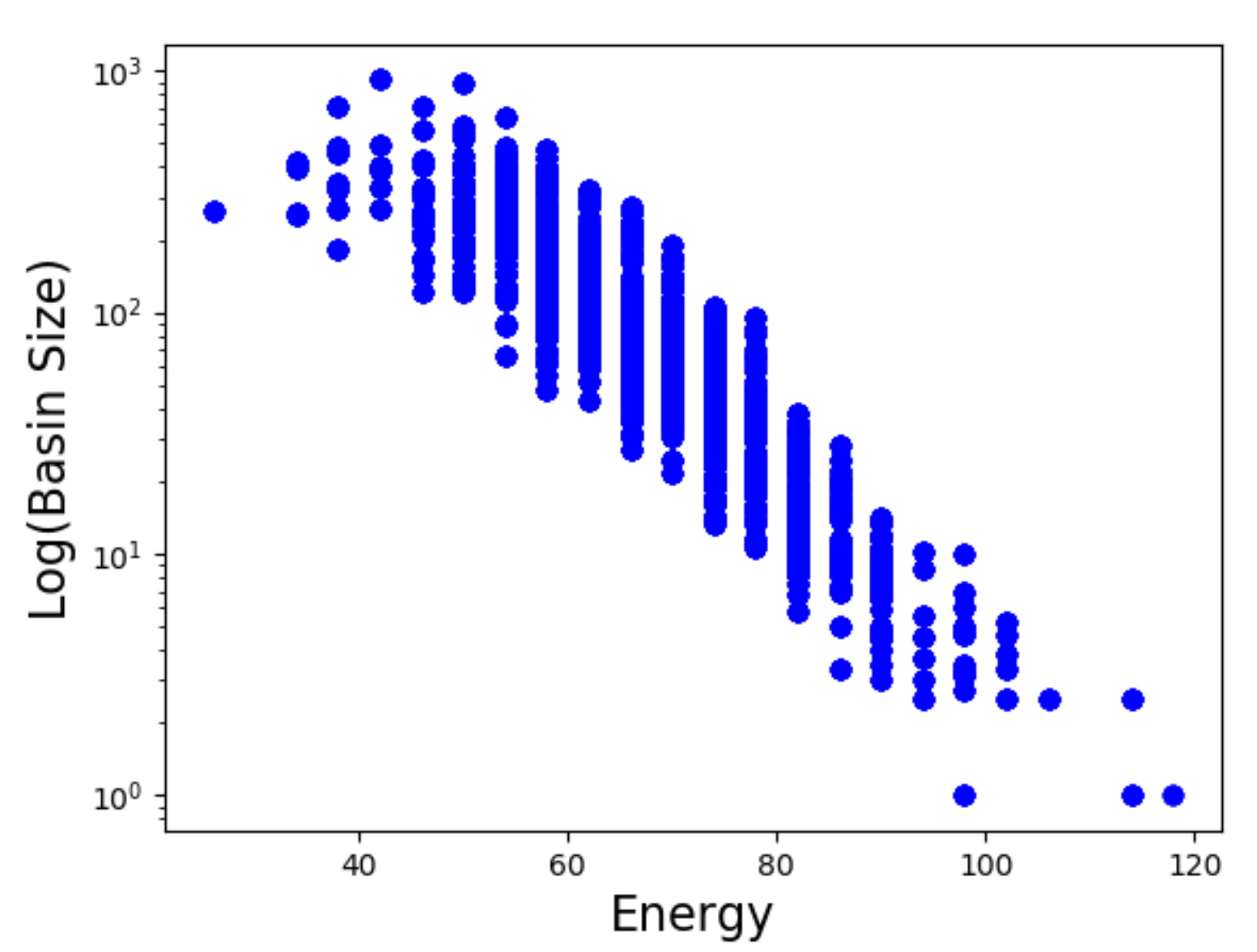} 
   \includegraphics[width=0.495\textwidth]{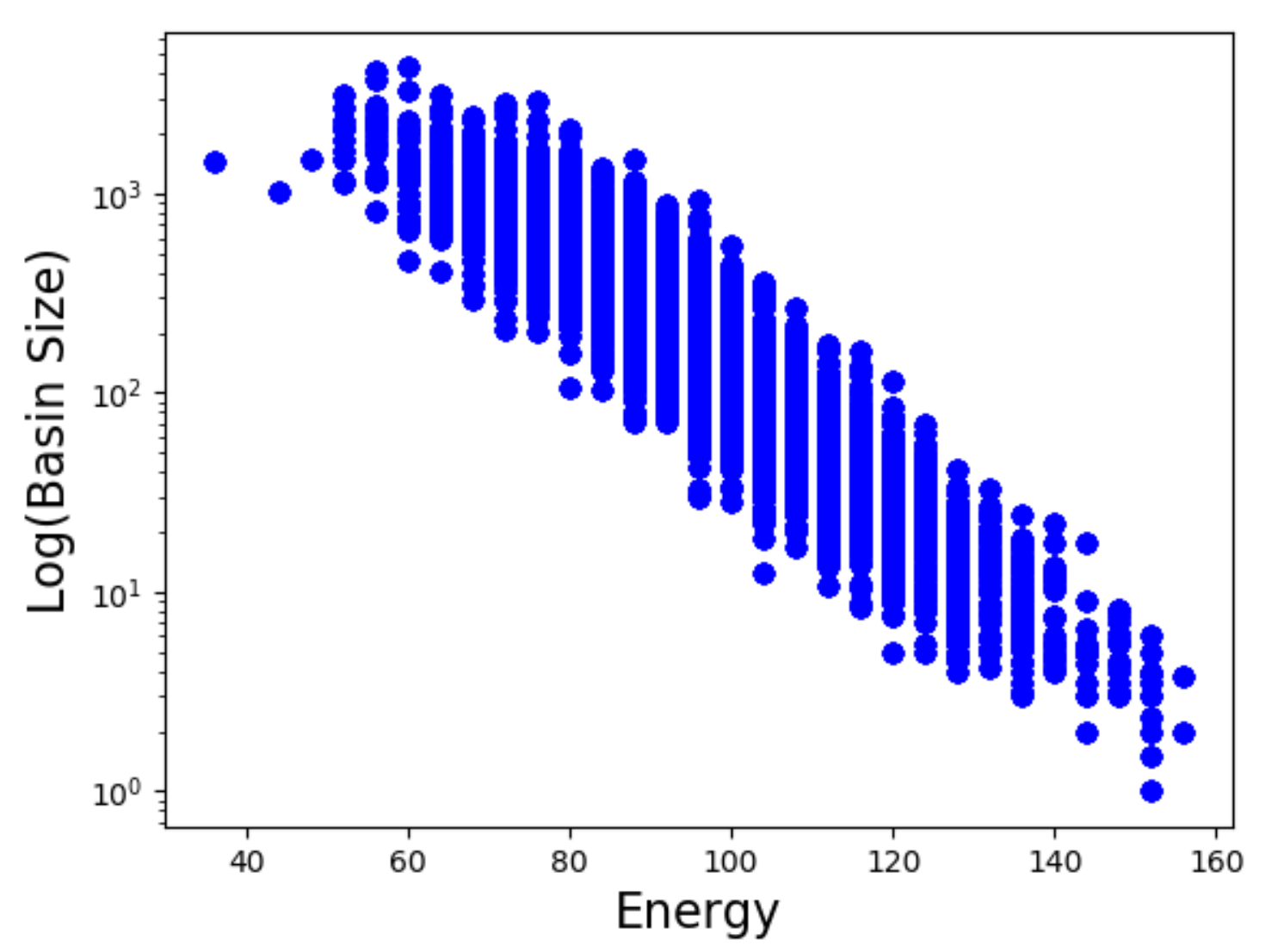} 
   \end{center}
      \vspace{-0.4cm}
   \footnotesize
    \hspace{2.8cm}$N=20   \hspace{5.2cm} N=24$
  \caption{Logarithm of the basin of attraction size of the minima in the LON as a function of their energy.
   Energy values are sorted in increasing order. Left image: LON corresponding to $N=20$; right image: $N=24$.}
\label{bas-en}
\end{figure*}

Figures~\ref{bas-en} show another aspect of the basin size distributions. The figures depict the basin size as a 
function of the corresponding minimum energy for $N=20$ (left image) and $N=24$ (right image). It clearly
appears that there is an inverse relationship between the energy of a given minimum and the size of the
corresponding basin. Thus, better minima have larger basin sizes. This is also in line with previous 
findings on
this and similar problems~\cite{ferreira2000,hernando2019}. Again, since the largest basin(s) correspond to the best energy minima, it
would appear that a search should be able to find one of them but actually this is not the case with $N$ large,
because there is a myriad of other basins with only slightly worse energy and similar size in which the 
search might get trapped. The previous considerations tend to invalidate the ``golf course'' hypothesis
put forward in~\cite{bernasconi} which states that the global minima for large $N$ are extremely rare
and are isolated in the energy landscape. In fact, they are indeed rare but
not isolated: in the picture that emerges examining their LONS, there is an exponentially
increasing number of less good minima everywhere around that may attract the search. A similar conclusion has
been reached by Ferreira et al.~\cite{ferreira2000}.

\subsection{Shortest Paths}

\begin{figure*}[h!]
  \begin{center}
   \includegraphics[width=0.6\textwidth]{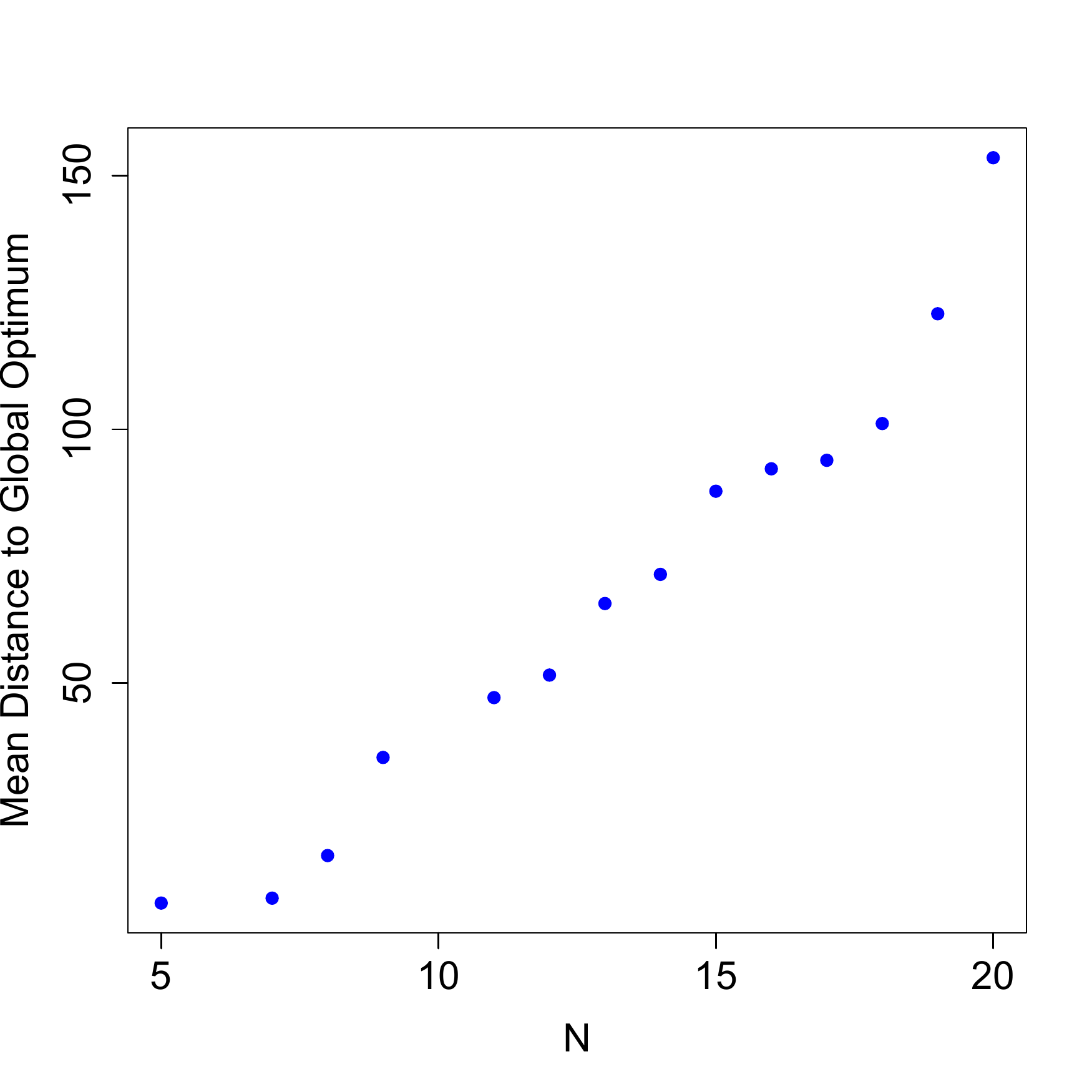}  
   \end{center}
  \caption{Average path lengths to a global optimum from all the other local optima for
  the LABS problems. Each point corresponds to a problem size $N$ for $N=5$ to $N=24$.
  The values $N=3,4,6,10$ are excluded because in this case all optima are global and thus the distance is zero.}
  \label{paths}
\end{figure*}

From the point of view of a search algorithm, a useful network measure is the average length of the shortest weighted path
from the local optima to the global ones. The edge weights are important because their values
are related to the transition probabilities between basins: an edge with a large weight means that
the corresponding transition is more likely to occur and thus the edge's length is proportional to
the inverse of the weight in the path computation using the customary Dijkstra algorithm. 
The averages are shown in Fig.~\ref{paths}. 
Only the distances for the values of $N$ that have non-degenerate sets of minima are shown since
for the others, i.e., for $N=3,4,6,10$ all the minima are global and thus the distance is zero. As well,
we limit $N$ to $20$ since the exponential increase of the number of minima makes the many source-single destination shortest
path algorithm to become too slow for larger $N$ values.

The implication is that the length of the path increases with problem size; in other words it becomes harder and
harder to reach a global optimum because the searcher must, on average, pass through an increasing number
of local optima before reaching the global one which, in turn, is a consequence of the exponential increase of the
number of optima. This can be alleviated to some extent by optimization heuristics such as simulated
annealing which can in principle overcome energy barriers in the high temperature regime.

\subsection{PageRank Centrality: Random Walks}

The PageRank~\cite{PR} algorithm provides a measure of the importance of a web page based
on how much it is referenced by other important pages. The aim is to capture the 
prestige of each node in order to rank pages by importance and thus reduce the amount of information
to process by an end user in a web search. The algorithm 
can be described as a random walk on the web considered as a directed graph. In a single step, the walker goes from a node
$a$ to a neighboring node $b$ with a probability given by $1/k$, where $k$ is the number of outgoing links from vertex $a$.
By definition of a random walk, after one step the probability distribution will be $\mathbf{ p}_1  = \mathbf{p}_0 \; \mathbf{T}$ and from step $j$ to $j+1$
it changes as 
\begin{equation}
\mathbf{ p}_{j+1}  = \mathbf{p}_j \; \mathbf{T}
\label{distr}
\end{equation}

\noindent Therefore, by iterating the previous equation from $j=0$ to $j=n-1$, the probability distribution ${\bf{ p}}_n$ after $n$ steps is:
$$
 \mathbf{ p}_n  = \mathbf{p}_0 \; \mathbf{ T}{^n},
$$

\noindent that is, it is given by  the initial probability distribution vector times the $n$-th power of the transition matrix 
$\mathbf{T}$. In the long time limit, if some conditions are satisfied, the probability distribution may reach an invariant value given by
 $\mathbf{p}^* = (p_1^*, \ldots, p_N^*)_{\infty}$ with $p_i^* = lim_{n \rightarrow \infty} p_{i}(n)$. 
Substituting in eq.~\ref{distr} gives
$$
 \mathbf{p}^*  = \mathbf{p}^* \mathbf{T}
 \label{lin-alg}
$$

\noindent From this eigenvalue equation the equilibrium, or stationary, probability distribution is the left eigenvector
of $\bf T$ with eigenvalue $1$~\cite{rand-walk}. This equilibrium probability distribution of the
above Markov chain
gives the asymptotic frequency of visits to each node of the network. In PageRank there is a device
that allows to get out of dead ends, i.e., when the walk reaches a node that does not have
outgoing links, or to simulate starting a new browsing session. In this case, with some probability
(around $0.2$) the walk jumps to a random node. This is not used here
because LONS are strongly connected graphs by construction.

\begin{figure*}[h!]
  \begin{center}
   \includegraphics[width=7cm, height=5cm, keepaspectratio]{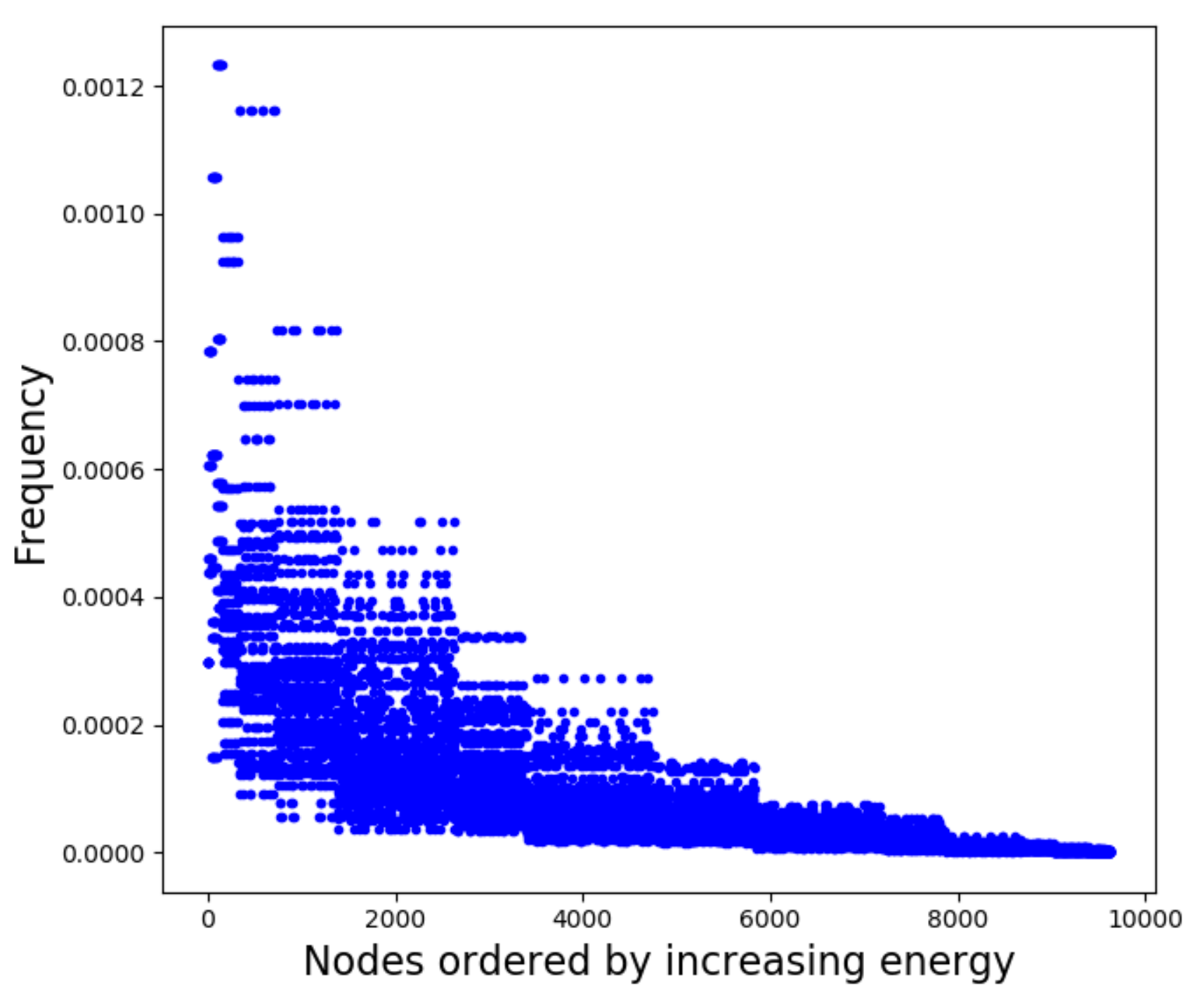}
   \includegraphics[width=7cm, height=5cm, keepaspectratio]{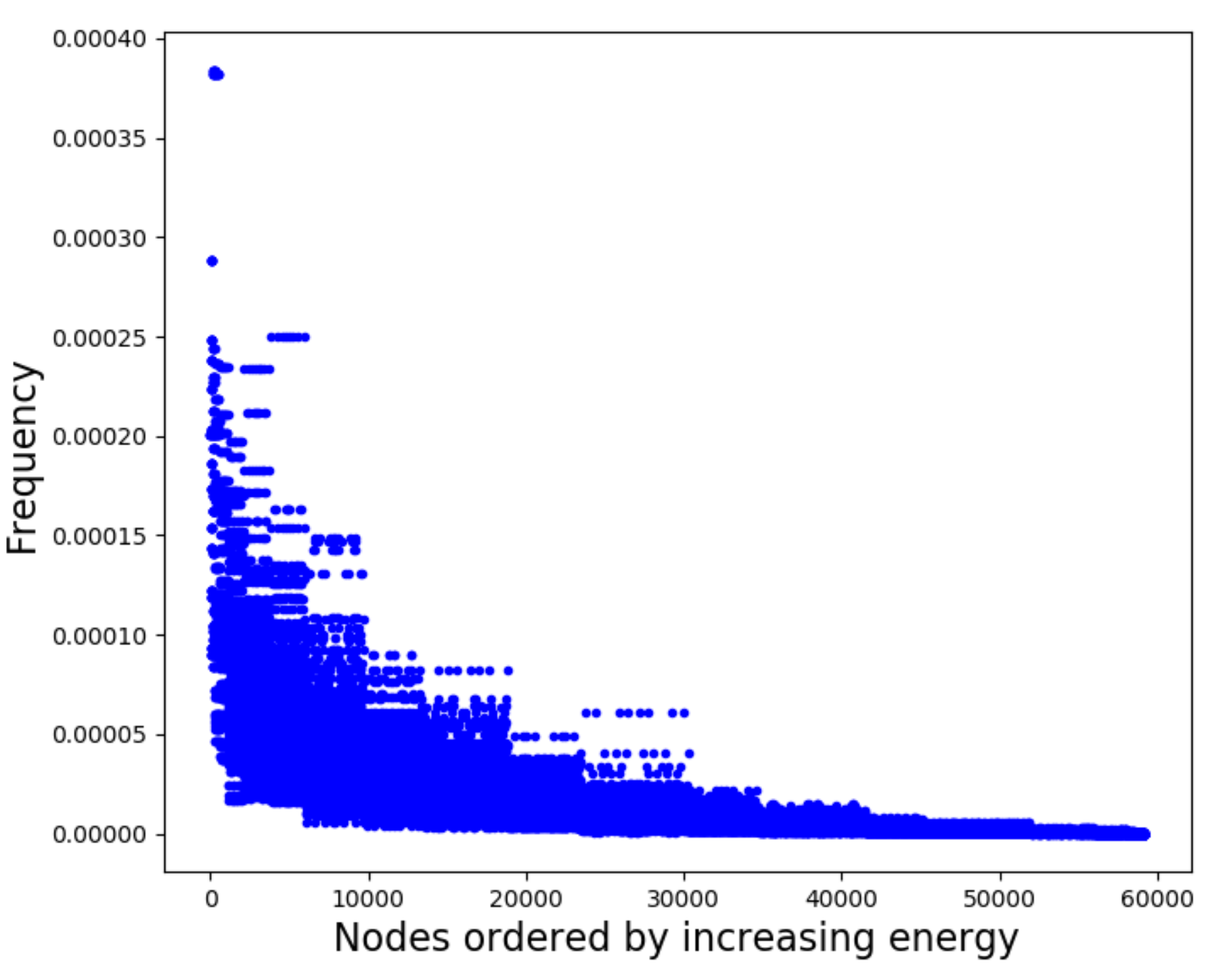} 
   \end{center}
         \vspace{-0.4cm}
   \footnotesize
   \hspace{2.8cm}$N=20   \hspace{5.5cm} N=24$
  \caption{PageRank centrality of all minima versus their energies as measured by the frequency of visits 
  during a random walk. Left picture: $N=20$; right picture: $N=24$. Energies increase to the right of 
  the x-axis.}
\label{pr}
\end{figure*}

PageRank has already been used in the context of problem difficulty in~\cite{herrmann2018}. 
Here we have applied PageRank to the LONS of the LABS problem. The results for $N=20$ and
$N= 24$ are given in Fig.~\ref{pr}, where nodes sorted by increasing energy are listed on the
x-axis, and the corresponding PageRank centrality, i.e.,  the long-term frequency
of visits of each node, is reported on the y-axis.

The important observation is that better optima, i.e. those close to the origin of the x-axis,
 are visited more often than less good ones. This is because the visit frequency depends
 on the incoming degree and, as better optima have larger basins, the corresponding LON
 nodes also have larger degree of the incoming links. Therefore, the result in terms of PageRank
 values is coherent with the picture that emerged previously when we analyzed the basins
 depths and sizes (see Fig.~\ref{bas-en}). Then, in principle, it would seem that the best optima should be easier
 to find than the worst ones. But this would be a wrong conclusion for two reasons. The first
 reason is that the ``searcher'' is just performing a random walk, in other words, it doesn't
 make use of the energy information associated to each LON node. In this way, the walk
 can only reach the global optimum by chance. The second reason is that with increasing $N$
 the number of optima increases exponentially (see Fig.~\ref{sizes-log}). As a result, there
 is a very large number of suboptimal, but still very good, minima that will be visited during the 
 random walk and only a handful among those are global optima.

\section{Conclusions}

In this contribution we have provided an up-to-date view of the structure of the energy landscape of the
LABS problem which has important applications in
communication engineering and in other fields, and is a typical representative
of the $NP$-hard class for which no polynomial algorithm is known. Since enumerative
algorithms cannot be used for large $N$ as they take too long to terminate, metaheuristics 
searches are often used. These searches
benefit the most from a better knowledge of the underlying energy landscape since they
work by sampling the latter in clever ways. To study the landscape features of interest we have made use 
of the \textit{local optima network} methodology by systematically extracting the problem optima
graphs for $N$ values up to $N=24$. Several metrics were used to characterize the
networks: number and type of optima, optima basins structure, degree and strength distributions,
shortests paths to the global optima, and random walk-based centrality of optima. 
Taken together, these metrics provide a quantitative and coherent explanation for
the difficulty of the LABS problem and give information about the energy landscape that
can be exploited for this problem as well as a number of other problems having a similar configuration
space structure. The methodology used here and the results presented above are also potentially 
useful for the investigation
of the energy landscape of spin glasses which, in spite of being disordered systems by definition,
have a number of similarities with the present problem, p-spin systems in particular. Future work is 
planned in this direction.

%\section{A Brief Comparison with a Simple Spin Glass}
%
%We have seen in the introduction that the LABS problem has many features in common with
%spin glass models with a major difference: interactions between ``spins'' are deterministic in LABS 
%while they are drawn from a probability distribution in spin glass models. Apart from this, LABS is
%formally similar to a four-spin system~\cite{bernasconi,ferreira2000} and it is possible to
%build a disordered model that can be viewed as the mean-field version of the 
%LABS~\cite{marinari1994,bouchaud1994}.
%
%Work is under way to investigate $p$-spin fitness landscapes with the tools introduced here for the LABS problem.
%However, for the time being, we conclude with 
%a brief discussion of differences and analogies that arise when comparing the LABS energy landscape
%with those of a very simple spin glass, the Sherrington-Kirkpatrick (SK) model~\cite{SK1975}.
%Assuming zero external magnetic field the SK Hamiltonian reads:
%$$
%H = -\frac{1}{\sqrt{N}} \sum_{i<k}J_{ik} s_i s_k,
%$$
%\noindent with the $J_{ik}$ drawn from a Gaussian distribution $\mathcal{N}(0,\sigma^2/N)$

\paragraph{Acknowledgements.} I am grateful to S\'ebastien V\'erel for useful discussions and for
producing the local optima graphs.

%\bibliographystyle{unsrt}

%\bibliography{lon}

\end{document}